# THE KEY EXCHANGE CRYPTOSYSTEM USED WITH HIGHER ORDER DIOPHANTINE EQUATIONS

Harry Yosh

HECO Ltd, Canberra, Australia
square17320508@yahoo.com

## ABSTRACT

*One-way functions are widely used for encrypting the secret in public key cryptography, although they are regarded as plausibly one-way but have not been proven so.*

*Here we discuss the public key cryptosystem based on the system of higher order Diophantine equations. In this system those Diophantine equations are used as public keys for sender and recipient, and both sender and recipient can obtain the shared secret through a trapdoor, while attackers must solve those Diophantine equations without trapdoor. Thus the scheme of this cryptosystem might be considered to represent a possible one-way function.*

*We also discuss the problem on implementation, which is caused from additional complexity necessary for constructing Diophantine equations in order to prevent from attacking by tamperers.*

## KEYWORDS

*public-key cryptography, one-way function, Diophantine equation, quotient ring, NP-complete*

## 1. INTRODUCTION

Since the public-key cryptography introduced by Diffie and Hellman,[1] various public-key cryptosystems based on integer factorization or discrete logarithm such as RSA, El Gamal, or those using elliptic curve techniques have been developed.[2] They obtain the computational security by using the sufficiently large key which imposes the attackers too heavy computational cost to decipher through solving integer factorization or discrete logarithm problems. Although no polynomial-time method for factoring large integers or solving the discrete logarithm problem has been known, it has not been proven that there exist no polynomial-time solutions for those problems.[3] Therefore the cryptosystems which are proven that there is no polynomial-time solution for attackers are thought to be significantly secure against their attacks.

Here a new cryptosystem is discussed. In this cryptosystem the public key for recipient is expressed with Diophantine equation and his/her private key is expressed with a value on the quotient ring defined with another Diophantine equation.

A Diophantine equation has plural unknown variables and defines an algebraic curve or algebraic surface. It is know to find the lattice points as the solutions of Diophantine equation is difficult in general when the order of that Diophantine equation is higher than one.[7] This difficulty is derived from that the equation may have no nontrivial solution, a finite number or an infinite number of solutions. Regarding its solvability, there is no general method to determine by a finite number of operations whether the equation is solvable or not.[8] However these properties of Diophantine equation are also useful for encrypting message to keep it secret and have been applied to some public key cryptosystems.[9][10] For example, the cryptosystem proposed by Lin, et al. is based on that the Diophantine equation dealt with in the system is practically non-soluble.[11] Precisely saying, the practically non-soluble Diophantine equation mentioned here indicates that there is no algorithm for solving it running in polynomial time. However it has been revealed that some alternative methods such as using AI[12], A* search[13] or genetic algorithm[14] are effective for solving numerically some specific types of Diophantine equation. It implies the cryptosystem built on a certain type of Diophantine equation might be broken within polynomial time by using those methods.

On the other hand, on the cryptosystem proposed here both sender and recipient can choose the form of Diophantine equation arbitrary. Therefore they can avoid the use of the specific types of Diophantine equation which are vulnerable to the methods mentioned above. The Diophantine equation used for recipient's public key is given by sender who has the trapdoor which is explained later and he/she can recovers the shared secret with that trapdoor, while attackers who have obtained the Diophantine equations as public keys for sender and recipient must solve them without the trapdoor. Thus this cryptosystem is expected to have the computational security stemmed from the difficulty of Diophantine problem.

In the following sections we discuss the basic idea for that cryptosystem, practical scheme and hardness of decryption by attackers. Also positive and negative aspects comparing with other cryptosystems are discussed.

## 2. KEY EXCHANGE SCHEME

Firstly we discuss the basic idea on this key exchange scheme using examples. To simplify the discussion we use simpler Diophantine equations and quotient ring in the following examples than those used in practical cases.

This key exchange scheme begins from the recipient side. The recipient sets the integer values of variables, say, x, y as x = 2, y = 3. Using those variables the recipient constructs a Diophantine equation such as,

$$x^3 - y^2 + 1 = 0 \qquad 1)$$

The recipient sends the information of that Diophantine equation to sender with keeping the values of x and y secret. The above Diophantine equation (1) is the public key and x, y are the private keys for recipient.

Sender constructs the polynomial as Diophantine equation on the quotient ring $Z[X,Y] / (X^3 - Y^2 + 1)$ through the following procedures; Firstly the sender defines the following operator on that quotient ring.

$$T_{[a,b;c]} : x \rightarrow (x + a)^c + b \qquad 2)$$

where a, b are integers and c is odd (c > 0). The inverse operator for it is given as,

$$T_{[a,b;c]}^{-1} : y \to (y - b)^{1/c} - a$$

3)

Sender then sets an element on the quotient ring, say $xy^2$, and applies the operator (2) repeatedly to that element as,

$$T_{[a1,b1;c1]} (T_{[a2,b2;c2]} (...(T_{[an,bn;cn]} (xy^2))...)) = T_{[a1,b1;c1]} T_{[a2,b2;c2]} ... T_{[an,bn;cn]} (xy^2)$$

$$= h(x,y)$$

4)

where $h(x,y)$ is an element in $Z[X,Y] / (X^3 - Y^2 + 1)$. For example,

$$T_{[1,2;3]} T_{[0,3;1]} (xy^2) = T_{[1,2;3]} (xy^2 + 3)$$

$$= (xy^2 + 4)^3 + 2 = x^3y^6 + 12x^2y^4 + 48xy^2 + 66$$

5)

Sender keeps the parameters in the operators (i.e. 1, 2, 3 and 0, 3, 1 in $T_{[1,2;3]}$ and $T_{[0,3;1]}$) secret. Those parameters are the private keys for sender and actually form a "trapdoor". $x^3y^6 + 12x^2y^4 + 48xy^2 + 66$ is expressed with various ways in $Z[X,Y] / (X^3 - Y^2 + 1)$. Sender chooses one of those representations, say, $(y^2 - 1) y^6 + 12x^2y^4 + 48xy^2 + 66 = y^8 - y^6 + 12x^2y^4 + 48xy^2 + 66$ and sends it with $xy^2$ to the recipient. The form of polynomial $y^8 - y^6 + 12x^2y^4 + 48xy^2 + 66$ is the public key for sender.

Recipient calculates $y^8 - y^6 + 12x^2y^4 + 48xy^2 + 66$ and $xy^2$ using his/her private keys $x = 2$, $y = 3$ as,

$$y^8 - y^6 + 12x^2y^4 + 48xy^2 + 66 = 10650$$

6)

$$xy^2 = 18$$

7)

and returns the value 10650 to sender with keeping $xy^2 = 18$ secret. The Diophantine equation (6) is another public key for recipient, and the value of $xy^2$ is actually the shared secret between sender and recipient as shown below; Sender calculates $xy^2$ using the inverse operators as,

$$T_{[0,3;1]}^{-1} T_{[1,2;3]}^{-1} (10650) = T_{[0,3;1]}^{-1} ((10650 - 2)^{1/3} - 1)$$

$$= T_{[0,3;1]}^{-1} (21) = (21 - 3) - 0 = 18$$

8)

and sender could recover the value of $xy^2$ as 18.

Attackers who have had the public keys for both sender and recipient must solve the following system of equations to obtain the value of $xy^2$.

International Journal of Network Security & Its Applications (IJNSA), Vol.3, No.2, March 2011

$$x^3 - y^2 + 1 = 0$$
9)

$$y^8 - y^6 + 12x^2y^4 + 48xy^2 + 66 = 10650$$
10)

Since the above system is zero-dimensional, the attackers can solve it numerically or using Groebner basis.[5] However when the number of variables are greater than two, or the system is positive-dimensional, solving it is difficult in general.[6]

Based on the above discussion, we construct the general scheme for the key exchange cryptosystem.

**Step 1**

Recipient sets the integer values of variables $x_1, x_2, \ldots, x_m$ as,

$$x_1 = k_1, x_2 = k_2, \ldots, x_m = k_m \qquad (k_j : \text{integers})$$
11)

and constructs Diophantine equation as recipient's public key using those variables as,

$$f(x_1, x_2, \ldots, x_m) = 0$$
12)

Recipient keeps $k_1, k_2, \ldots, k_m$ secret.

**Step 2**

Recipient sends the above Diophantine equation (12) to sender.

**Step 3**

Sender sets an element $g(x_1, x_2, \ldots, x_m)$ in the quotient ring $Z[X_1, X_2, \ldots, X_m] / (f(X_1, X_2, \ldots, X_m))$ and defines the operators $T_{[a_j, b_j : c_j]}$ ($j = 1, \ldots, n$) on that quotient ring as,

$$T_{[a_j, b_j : c_j]} : x \rightarrow (x + a_j)^{c_j} + b_j$$
13)

where $a_j, b_j$ are integers and $c_j$ is odd ($c_j > 0$). Then the sender applies those operators to $g(x_1, x_2, \ldots, x_m)$ as,

$$T_{[a_1, b_1 : c_1]} (T_{[a_2, b_2 : c_2]} ( \ldots (T_{[a_n, b_n : c_n]} (g(x_1, x_2, \ldots, x_m))) \ldots )$$

$$= T_{[a_1, b_1 : c_1]} T_{[a_2, b_2 : c_2]} \ldots T_{[a_n, b_n : c_n]} (g(x_1, x_2, \ldots, x_m))$$

$$= h(x_1, x_2, \ldots, x_m)$$
14)

where $h(x_1, x_2, \ldots, x_m)$ is an element in $Z[X_1, X_2, \ldots, X_m] / (f(X_1, X_2, \ldots, X_m))$ and has generally various representations. Sender chooses one of them as his/her public key.

**Step 4**

Sender sends the public key $h(x_1, x_2, \ldots, x_m)$ chosen in the previous step with the form $g(x_1, x_2, \ldots, x_m)$ to recipient.

**Step 5**

Recipient calculates $h(x_1, x_2, \ldots, x_m) = p$ and $g(x_1, x_2, \ldots, x_m) = s$ using (11) where p and s are some integers.

**Step 6**

Recipient sends back the value p to sender with keeping the value s secret.

**Step 7**

Sender recovers the value s of $g(x_1, x_2, \ldots, x_m)$ as,

$$s = T_{[an,bn:cn]}^{-1} \, T_{[an-1,bn-1:cn-1]}^{-1} \ldots T_{[a1,b1:c1]}^{-1} (p) \quad (15)$$

Thus sender and recipient could share the secret s.

The parameter c in the operator $T_{[a,b;c]}$ must be odd, otherwise the image of inverse operator;

$$T_{[a,b;c]}^{-1} : y \to (y - b)^{1/c} - a \quad (16)$$

may not be determined uniquely and it causes the ambiguity on the obtained secret s.

It is possible to modify the above scheme to that on finite ring. In that case the Diophantine equation (12) is modified to a congruence equation with modulus w and the quotient ring is rewritten as,

$$Z_w[X_1, X_2, \ldots, X_m] / (f(X_1, X_2, \ldots, X_m)) \quad (17)$$

where $Z_w$ is the quotient ring composed of 0, 1, ..., w-1. The operator defined in Step 3 is modified as,

$$T_{[a,b:c]} : x \to (x + a)^c + b \quad \mod w \quad (18)$$

and its inverse operator is expressed as,

$$T_{[a,b;c]}^{-1} : y \to (y - b)^{c'} - a \mod w$$

19)

where $cc' = 1 \mod \varphi(w)$ ($\varphi(w)$: Euler's totient function).

## 3. HARDNESS OF DECRYPTION FOR ATTACKERS

Attackers who have had the public information of sender and recipient are to solve the following system of two Diophantine equations;

$$f(x_1, x_2, ... , x_m) = 0 \quad ... \text{ in Step 1}$$

20)

$$h(x_1, x_2, ... , x_m) = p \quad ... \text{ in Step 5}$$

21)

and estimate the secret $s = g(x_1, x_2, ... , x_m)$ with the obtained solution. As mentioned in earlier section, it is difficult to solve (20) and (21) in integers when m is greater than two in general.

As another strategy, attackers may consider to decipher the sequence of operators appeared in Step 3;

$$T_{[a1,b1:c1]} T_{[a2,b2:c2]} ... T_{[an,bn:cn]} (g(x_1, x_2, ... , x_m)) = h(x_1, x_2, ... , x_m)$$

22)

by using (20) and (21). Once the above sequence has been deciphered, attackers can estimate the secret $s = g(x_1, x_2, ... , x_m)$ applying the inverse sequence of the above sequence. However the highest order among $x_1, x_2, ... , x_m$ in $h(x_1, x_2, ... , x_m)$ becomes greater in general along with the increase of the iteration number n expressed in (22). Since the greater the highest order is, the more equivalent expressions in the quotient ring $Z[X_1, X_2, ... , X_m] / (f(X_1, X_2, ... , X_m))$ are, it is also virtually impossible to decipher the sequence of operators (22) unless attackers have known the iteration number n previously or the Diophantine equation (12) is linear.

When the system of higher order congruence equations are used for the scheme instead of Diophantine equations (20) and (21) as mentioned in earlier section, the size of secret key is generally determined by that of modulus w. Although the solution space for the system of congruence equations is finite, obtaining the secret s is still difficult for attackers because in general the problem to solve such system of higher order congruence equations is NP-complete when m > 2.[4]

## 4. PRO AND CON

Attackers must solve the system of Diophantine equations to estimate the shared secret between sender and recipient. Generally it is hard to solve that system when it is positive-dimensional. Attackers also subject the similar constraint on the corresponding higher order congruence equation system. That feature brings the proven security which does not depend on the performance of attackers' computational capacity. Namely sender and recipient can choose the public keys which have the length shorter than that of public keys used for other cryptosystems without harming the security, and the shorter length of public keys would alleviate the computational load considerably.

Comparing with other key exchange cryptosystem such as RSA or Diffie-Hellman scheme, this cryptosystem needs additional key exchange step, that is, recipient must tell a Diophantine equation from which the quotient ring is constructed to sender prior to key exchange between them. It may increase the vulnerability regarding the security on this key exchange cryptosystem.

Sender and recipient exchange Diophantine equations as their public keys. Although sender and recipient can choose their Diophantine equations arbitrary, some specific considerations are required to construct those Diophantine equations practically in the actual key exchange scheme as discussed below.

Firstly in Step 1 described in the section: Key exchange scheme, the Diophantine equation (12) must be constructed carefully to avoid the case that it has unique solution, otherwise as soon as solving that equation, attackers get the secret keys expressed in (11). Next in Step 3, the Diophantine equation (14) must not have unique solution as well as (12), also it must be complicated enough to prevent from inferring the sequence of operators used for constructing (14) by attackers. It is not practical that the above Diophantine equations are constructed manually by sender and recipient in general cases, and it will be necessary to implement appropriate programs into that cryptosystem to generate those equations automatically based on the values of variables $x_1$, $x_2$, ... , $x_m$ set in Step 1, and $a_j$, $b_j$, $c_j$ and n set in Step 3. To prevent from breaking the patterns of generating them by attackers, the forms of Diophantine equations determined by those programs must not be static, but dynamically change according with the input, or randomly. Thus the implementation of this cryptosystem will impose additional calculation load for them with that for dealing with the Diophantine equations symbolically.

## 5. SUMMARY

In the proposed key exchange cryptosystem the system of two higher order Diophantine equations is considered for encrypting shared secret between sender and recipient.

The first Diophantine equation is created by recipient and sender creates another Diophantine equation on the quotient ring based on the first Diophantine equation. The shared secret has the form of polynomial and both sender and recipient can obtain its value by a trapdoor without solving the system of those Diophantine equations explicitly.

On the other hand, attackers must solve the following system of two Diophantine equations mentioned above in order to obtain the secret.

$$f(x_1, x_2, ... , x_m) = 0$$

23)

$$h(x_1, x_2, ... , x_m) = p$$

24)

In general it is hard to solve the above system in integers for $x_1$, $x_2$, ... , $x_m$ when m is greater than two.

Although this key exchange cryptosystem has the intrinsic security mentioned above, it requires rather complicated implementation comparing with other key exchange cryptosystems.

This complexity is due to generate Diophantine equations used as public keys with unpredictable manner lest attackers detect the pattern of their generations. The specifically designed program must be implemented in the cryptosystem for that purpose.

## REFERENCES

[1] W. Diffie and M. E. Hellman, *New Directions in Cryptography.* IEEE Transactions on Information Theory, vol. IT-22, pp. 644-654, Nov. 1976.

[2] D. Knuth, *The Art of Computer Programming, Volume 2: Seminumerical Algorithms, Third Edition.* Addison-Wesley, 1997.

[3] M. Sipser, *Introduction to the Theory of Computation.* PWS Publishing, pp. 374-376, 1997.

[4] M. R. Garey, D. S. Johnson, *Computers and Intractability: A Guide to the Theory of NP-Completeness.* W. H. Freeman, 1979.

[5] B. Buchberger, *An algorithmic criterion for the solvability of algebraic systems of equations.* Aequationes Math., 4, pp. 374-383, 1965.

[6] Yu. I. Manin, *Hilbert's tenth problem.* J. Soviet Math., 3 : 1, pp. 164-184, 1975.

[7] H. Cohen, *Number Theory, Vol. I: Tools and Diophantine Equations and Vol. II: Analytic and Modern Tools.* Springer-Verlag, GTM 239, 240, 2007.

[8] Y. V. Matiyasevich, *Hilbert's Tenth Problem.* MIT press, 1993.

[9] C. S. Laih, et al, *Cryptanalysis of Diophantine equation oriented public key cryptosystem.* IEEE Transactions on Computers, Vol 46, April, 1997.

[10] H. Ong, C. Schnorr, A. Shamir, *An efficient signature scheme based on polynomial equations.* Proc. of CRYPTO, pp 37–46, 1985.

[11] C. H. Lin, C. C. Chang, R. C. T. Lee, *A New Public-Key Cipher System Based Upon the Diophantine Equations.* IEEE Transactions on Computers, Volume 44, Issue 1, 1995.

[12] G. Luger, *Artificial Intelligence: Structures and Strategies for Complex Problem solving.* 4e Pearson Education, 2006.

[13] S. Abraham, and M. Sanglikar, *A\* Search for a Challenging Problem.* Proceedings of the 3rd International Conference on Mathematics and Computer Science held at Loyola College, Chennai, 5th- 6th January 2009.

[14] Z. Michalewich, *GA + Data Structures = Evaluation Programs.* Springer Verlag, 1992

**Author**

Harry Yosh is a knowledge manager at HECO Ltd. He is also a data analyst at APA Group, Canberra, having graduated from University of Canberra with a Master of IT. His hobbies are traveling and swimming.

Paper: Harry Yosh, *General Addition Formula for Meromorphic Functions Derived from Residue Theorem.* Journal of Mathematical Physics, Analysis, Geometry, v. 6, No 2, p. 183-191, 2010. (http://jmage.ilt.kharkov.ua/abstract.php?uid=jm06-0183e)